# Half-Metallicity of Wurtzite NiO and ZnO/NiO (0001) Interface: First Principles Simulation

Z. P. Chen[1], L. Miao[1,†] and X. S. Miao[1,2]

[1] *Department of Electronic Science and Technology, Huazhong University of Science and Technology, Wuhan, HUBEI 430074, China*

[2] *Wuhan National Laboratory for Optoelectronics, Wuhan, HUBEI 430074, China*

First principles calculations based on density functional theory are performed to investigate the structural, electronic and magnetic properties of wurtzite ZnO/NiO (0001) interface. By using DFT+$U$ method we discover that the half-metallic behavior of wurtzite NiO ($w$-NiO) retains in the ZnO/NiO (0001) interface. Through analyses of density of state, charge population and magnetic moments, we find the half-metallicity is weakened around the interface but interface effect is quite localized. More over the interface system keeps a ferromagnetic ground state as bulk $w$-NiO does. Based on the simulations of epitaxial growth case, $w$-NiO is predicted to be a promising candidate of electrode for the injection of spin polarized currents.

PACS number(s): 73.20.At, 75.70.Cn, 85.75.－d

## Ⅰ. INTRODUCTION

Injection of spin polarized electrons － the key factor for achieving spintronics － has drawn great attention during recent decades.[1, 2] One popular way is using ferromagnetic (FM) metal as an injection source of spin polarized electrons.[3, 4] However, one problem revealed is that due to the conductivity mismatch between FM metal and semiconductor, the measured spin polarized current is rather weak.[4, 5] Until now many magnetic materials have been discovered to have half-metallic property,[6-8] which exhibits a metallic density of state (DOS) in one spin channel but a band gap around the Fermi level in the other.[6] With 100% spin polarization ideally, half-metal is a better spin injection source than FM metals.[2] So far, there have been many attempts to use half-metal in spintronic devices, such as FM electrodes of spin valves[9] or magnetic tunnel junctions.[10, 11] Nevertheless first principles calculations have shown that half-metal/semiconductor interface configuration influence the electronic properties and in turn may destroy the spin polarization of the carriers.[12-14]

---

[†] E-mail: miaoling@mail.hust.edu.cn

Wurtzite NiO ($w$-NiO) has been found to display half-metallicity.[15] Also $w$-NiO was found to have a small lattice mismatch with those most popular wurtzite wide gap semiconductors such as ZnO, GaN, and SiC.[15] So even though wurtzite structure is a metastable phase for NiO, its stability may be achieved by epitaxially growth on those semiconductor substrates. Then if half-metallicity of $w$-NiO is not destroyed at the interface, injection of 100% spin polarized electrons will be achieved. Among those wide gap semiconductors ZnO is the most promising candidate of substrate material since the common O element of both ZnO and NiO enhances the possibility of successful epitaxial growth of this kind. Also the interface O atomic monolayer make Zn and Ni atoms separated and their interaction screened to some degree. So this interface configuration may have a relatively small effect on the electronic properties and thus $w$-NiO will keep half-metallicity.

Recently, Masuko et al.[16, 17] have took several experiments on spin-dependent transport of ZnMnO/ZnO heterostructure. They used single crystal ZnO with atomically smooth surface as substrate and deposit ZnMnO thin film on it using pulsed laser deposition method. Via magnetoresistance measurements, they studied spin-dependent transport properties of the heterostructure and confirmed the spin polarization at the (0001) interface. Analogously, if ZnO/$w$-NiO heterostructure can be experimentally fabricated, the same method can be utilized to investigate the spin polarization of $w$-NiO and to confirm whether the injection of spin polarized electrons is feasible. Nevertheless, neither experimental nor theoretical investigations of spin polarization of *w*-NiO/semiconductor interface has been carried out so far.

In this paper, first-principles calculations have been performed on bulk *w*-NiO and ZnO/NiO (0001) interface based on density functional theory (DFT). Discussions of results have been focused on the structural, electronic and magnetic properties. We will show that *w*-NiO remain half-metallic at the interface against ZnO and it has promising applications in spintronics.

## Ⅱ. COMPUTATIONAL DETAILS

We employ the projector augmented wave (PAW)[18, 19] method in the framework of spin polarized density functional theory (DFT)[20, 21] as implemented in the Vienna *ab initio* simulation package (VASP)[22, 23]. The generalized gradient approximation (GGA)[24] for the correlation functional is used in the calculation with an energy cut off of 400 eV. The Brillouin zone is sampled by Γ centered 4×4×5, and 6×6×2 grid using the Monkhorst-Pack scheme[25] for bulk and interface structures, respectively. The structure relaxation is performed using the conjugate gradient method[26] and is done when the force acting on each atom is less than 0.05 eV/Å.

The ZnO/NiO (0001) interface is constructed applying (2×2×4) slab supercell with 64 atoms consisting of 8 ZnO monolayers and 8 *w*-NiO monolayers with a layer sequence of …$Zn^{-3}/O^{-2}/Zn^{-1}/O^{0}/Ni^{1}/O^{2}/Ni^{3}/O^{4}$…. The superscripts indicate the relative positions of the monolayers along *c* axis; cf. FIG. 1. Noteworthily, there are two types of interfaces here and both of them probably exist in experimental situations. In interface Ⅰ type, the communal $O^0$ layer is closer to $Ni^1$ and one $O^0$ atom bond with three $Ni^1$ neighbors but one $Zn^{-1}$ neighbor. On the other hand, at interface Ⅱ, $O^8$ atoms bond with $Ni^7$ and $Zn^{-7}$ atoms forming three Zn-O bonds against one Ni-O bond as shown in FIG. 1.

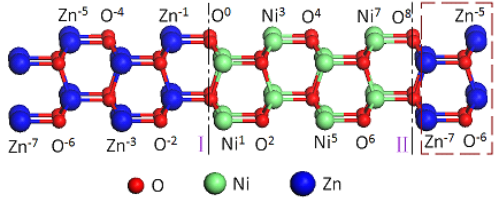

FIG. 1 (color online). Atomic configuration of (2×2×4) ZnO/NiO (0001) interface structure. The dash dot lines with notations indicate the O layers of interfaces Ⅰ and Ⅱ. The atoms in the dashed box are repetition of the left-most three atomic layers due to periodic boundary condition.

For transition metal semicore $3d$ states, DFT is known to have an inadequate description and the calculated binding energy and electronic properties deviate from the experimental results.[27, 28] For the correction of pure DFT we adopt the DFT+$U$ approach.[28, 29] The value used by P. Gopal *et al.*[30] is adopted here choosing $U$ = 4.5 eV and $J$ = 0.5 eV on the semicore $d$ states of both Zn and Ni ions. It has been demonstrated that a small value of $U$ like this can give a more accurate description of $d$ electrons of transition metal ions.[19, 31, 32]

## Ⅲ. RESULTS AND DISCUSSION

We first examine the structural and electronic properties of bulk ZnO, and compare the results with other theoretical and experimental data. The resulting lattice constant is $a$ = 0.321 nm, $c$ = 0.516 nm, which is in good agreement with experimental value of $a$ = 0.325 nm, $c$ = 0.521 nm.[33] The band gap obtained is $E_g$ = 1.45 eV, which is similar to the result of $E_g$ = 1.51 eV with $U$ = 4.7 eV calculated by A. Janotti *et al.*[34] Although both results of $E_g$ deviate considerably from the experimental result ($E_g$ = 3.43 eV),[33] the DFT+$U$ results give better descriptions of the ZnO electronic structure than the pure DFT method ($E_g$ = 0.8 eV)[34].

Next we calculated the properties of bulk $w$-NiO. The fully relaxed lattice constant is $a = 0.297$ nm, $c = 0.545$ nm. Consistent with the discovery of Wu *et al.*[15], for the majority spin channel the Fermi level located in the band gap, and for the minority spin there is no energy gap around the Fermi level, which gives a direct evidence of the half-metallic behavior of $w$-NiO. The calculated band gap is $E_g = 1.582$ eV, and the spin-filp gap $E_g^{sp} = 1.263$ eV, which is defined as energy difference between the Fermi level of the metallic spin and the conduction band minimum of the semiconducting spin.[15] The ground state of $w$-NiO is found to be a ferromagnetic state with a magnetic moment of 1.666 $\mu_B$ on the Ni atom and an induced moment of 0.275 $\mu_B$ on the O atom. The lattice mismatch of in-plane lattice constant $a$ between $w$-NiO and ZnO is about 7.5%. In comparison we perform another calculation of $w$-NiO under a biaxial tensile strain. In this case the in-plane lattice constant $a$ of $w$-NiO is fixed to that of ZnO (0.321 nm), while the lattice constant $c$ and the atomic positions are fully relaxed. Still, $w$-NiO presents half-metallic behavior in this constrained case. In comparison with the previous case the band gap $E_g = 1.141$ eV and the spin-filp gap $E_g^{sp} = 0.811$ eV is narrower. The ferromagnetic ground state remains with a magnetic moment of 1.640 $\mu_B$ on the Ni atom and 0.318 $\mu_B$ on the O atom.

As mentioned previously, the metastable $w$-NiO will possibly keep stable with wurtzite structure by practical approach of epitaxial growth on ZnO substrate. To discover the interface effects acting on w-NiO, we perform calculations of ZnO/NiO (0001) interface and investigate corresponding electronic and magnetic properties. Since NiO suffers biaxial strain in epitaxial growth situation we first set the in-plane lattice constant $a$ to that of calculated ZnO; meanwhile the lattice constant $c$ and the atomic positions are fully relaxed for minimization of total energy.

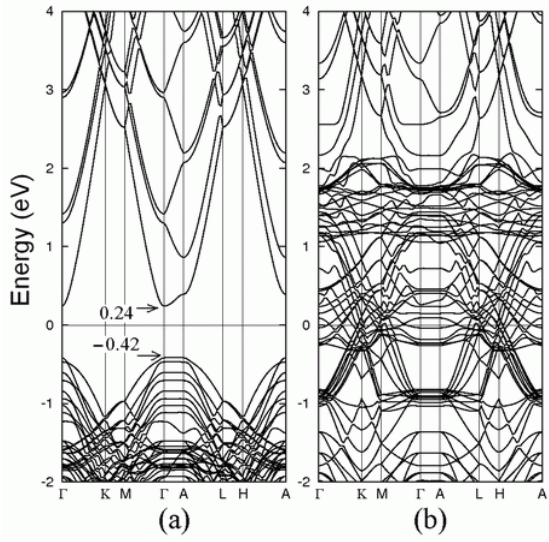

FIG. 2. (a) and (b): band structure of majority spin (up) and minority spin (down) of ZnO/NiO (0001) interface with in-plane lattice constant $a$ set to that of ZnO. The Fermi level is set to zero.

FIG. 2 shows the spin-resolved band structure of ZnO/$w$-NiO (0001) interface. This structure exhibits the desired half-metallic behavior with a band gap of ~0.66 eV in the majority spin channel and with the Fermi level across the valance band in the minority spin channel. The spin-flip gap is $E_g^{sp}$ = 0.24 eV. Both $E_g$ and $E_g^{sp}$ become narrower in comparison with those of bulk $w$-NiO mentioned above. The main reason for this gap reduction is the overlap of ZnO band and NiO band. The conduction band minimum (~0.24 eV above the Fermi level) of majority spin consists of ZnO states. While the valance band maximum (~0.4 eV below the Fermilevel) of majority spin is formed by states of NiO. This can be demonstrated by spin-dependent atomic projected density of state; cf. FIG. 3. So the majority band gap (~0.66 eV) of interface system is the energy difference between ZnO conduction band minimum and NiO valance band maximum.

Noteworthily, despite the band gap reduction caused by band overlap, the majority spin energy gap of bulk $w$-NiO (1.141 eV) and the band gap of bulk ZnO (1.45 eV) change very little in interface system according to FIG. 3. In comparison with DOS of bulk ZnO (not shown here) the uniform Fermi level in FIG. 3 has been lifted up for 1.2 eV from valance band maximum of ZnO, but it remains in the energy gap. In contrast, the uniform Fermi level has moved downwards from the original position in the majority gap of bulk $w$-NiO. Significantly, the Fermi level do not move into the majority spin valance band of $w$-NiO, and thus the interface system exhibits half-metallicity as bulk $w$-NiO does.

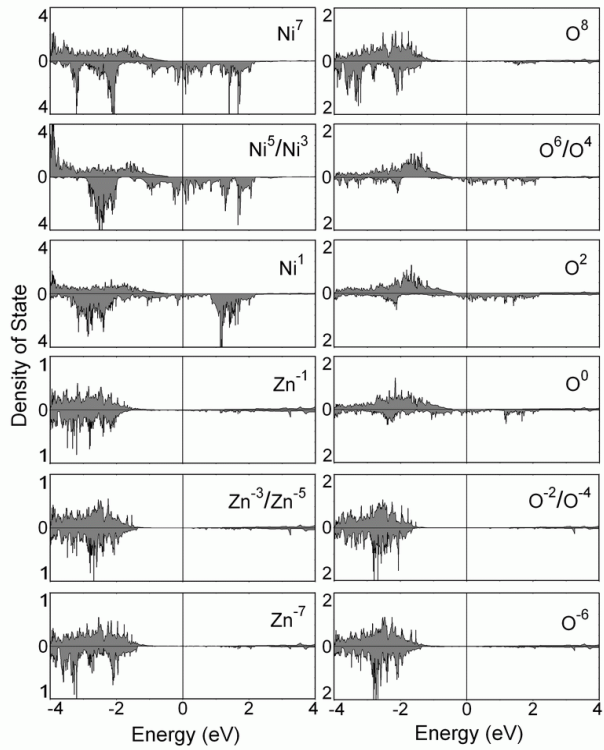

FIG. 3. Spin-dependent atomic PDOS of the ZnO/NiO (0001) interface. The notations indicate the atomic kinds and the relative positions of corresponding monolayer. The upper part of each plot represents the majority spin (up), vise versa. Atoms of some inner neighboring layers have almost identical PDOS and they are plotted in the same graph. The Fermi level is set to zero.

The influence brought by interface to the electronic properties of both $w$-NiO and ZnO is found to be localized. Unlike interfacial atoms, all those inner atoms show the same PDOS behavior as corresponding bulk materials do; cf. FIG. 3. Only atoms near the two interfaces shown in FIG. 1 present different PDOS. At interface Ⅰ, half-metallicity of $Ni^1$ and $O^0$ become weaker as their DOS of minority spin decrease around Fermi level. Since $O^0$ is closer to $w$-NiO part its DOS is more similar to $w$-NiO ($O^2$) than to ZnO ($O^{-2}$), and $Ni^1$ is more affected than $Zn^{-1}$. In contrast, at interface Ⅱ, $Ni^7$ is away from $O^8$ so the half-metallic behavior is as strong as the inner ones ($Ni^3$, $Ni^5$). While $O^8$ becomes totally insulate as ZnO ($O^{-6}$) does.

Evidence of the localization of interface effect can also be found from charge population of atoms at each monolayer; cf. FIG. 4. Inner atoms keep similar charge value to the atoms of bulk material. At interface Ⅰ, O charge holds the intermediate value of both sides; while at interface Ⅱ O charge keeps the same value as ZnO. Ni charge decrease at both two interfaces, but Zn charge increased rather slightly. This charge transfer indicates these interface atoms denote electrons to the $p$-$d$ bonding bands. As an evidence, we find the $p$ states of interface $O^8$ hybridize with neighboring $Ni^7$ $d$ states and $Zn^{-7}$ $p$, $d$ states contributing to the two minority spin PDOS peaks at around $-2.1$ eV and $-3.2$ eV; cf. FIG. 3.

Changes of atomic magnetic moments are obvious merely around interfaces, since inner O atoms and Ni atoms keep magnetic moments close to those of bulk materials, respectively. At interface Ⅰ and Ⅱ, O moments value are between the O moments of inner ZnO and inner NiO, and Ni moments increase at the two interfaces. Zn moments still keep around zero like bulk ZnO does. Noteworthily, the difference between magnetic moments of interfacial $Ni^1$ and inner $Ni^3$ is about 0.08 $\mu_B$, very close to the charge difference of ~0.08 e between corresponding Ni atoms; and majority spin PDOS of interfacial $Ni^1$ and $Ni^7$ have changed little compared to inner ones; cf. FIG. 3. This means only minority spin (down) electrons transfer from interfacial Ni atoms and accordingly contribute to the increase of magnetic moments. In contrast, between interfacial and inner O atoms, differences of magnetic moments are a magnitude larger than differences of charges. For interfacial O, half-metallicity is weakened since some minority states originally around the Fermi level are pushed downwards to the valance band; cf. FIG. 3. So some O electrons flip from the spin up channel to the spin down channel making interfacial O moments decrease considerably. All Ni, O, Zn atoms have positive magnetic moments, indicating the ferromagnetic ground state of the interface system, in accordance with the *w*-NiO character.

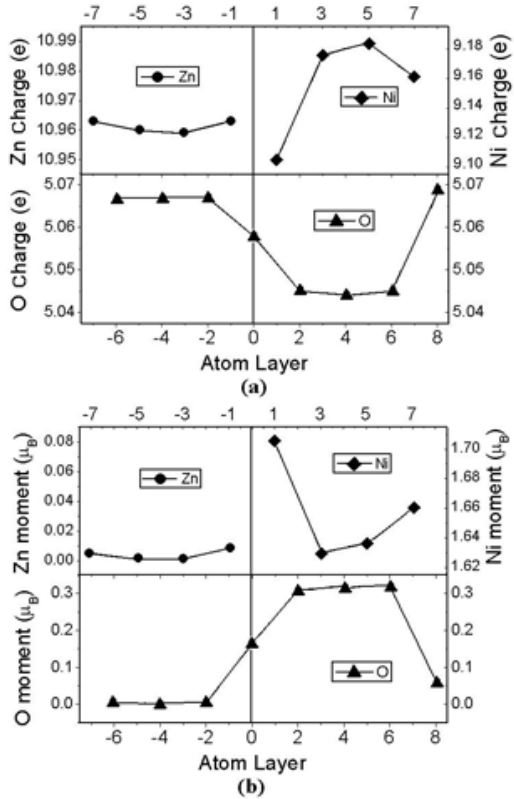

FIG. 4. Charge population and magnetic moments of Ni, O and Zn atoms at different monolayers. Abscissa indicates the relative positions of the monolayers along *c* axis.

Under some other experimental conditions rather than epitaxial growth, ZnO do not necessarily act as substrate when NiO and ZnO form a heterostructre, and we can expect that in these cases both ZnO and NiO will suffer from a strain at the interface because of the lattice mismatch. For instance, Meyerheim *et al*.[35] have found that wurtzite type CoO nanocrystals were embedded in Co doped ZnO thin films and distortion existed at the interface. For this reason, we perform calculations in the case where all the lattice constants and atomic positions are fully relaxed, and discover how the biaxial strain will influence the structural, electronic and magnetic properties. The resulting in-plane lattice constant a = 0.312 nm, is 2.8% smaller than that of calculated ZnO and 5.1% larger than that of *w*-NiO. Both ZnO and NiO are found to be mutually strained at the interface, as expected. Most significantly, the ZnO/NiO (0001) interface system is still half-metallic in this case. The calculated majority spin band gap $E_g$ = 0.911 eV and spin-flip gap $E_g^{sp}$ = 0.487 eV are narrower than that of bulk *w*-NiO but broader than that of former one in fixed lattice constant *a* case. In spite of the band gap difference, the ZnO/NiO (0001) interface system in this fully relaxed lattice case exhibit similar PDOS and magnetic behavior to the case as discussed above.

So far, several experimental studies have been performed focusing on spin polarization of the interface between magnetic materials and ZnO, such as $Mg_xZn_{1-x}O$/ZnO,[36] ZnMnO/ZnO[16, 17] and CoO/ZnO[35]. Different methods of sample preparation and measurements have been utilized in these experiments. In order to confirm the half-metallicity or spin polarization of *w*-NiO or ZnO/NiO (0001) interface, corresponding experiments must be taken using those methods.

## Ⅳ. CONCLUSION

Based on the density functional calculations, we investigate the structural, electronic and magnetic properties of bulk *w*-NiO and ZnO/NiO interface both in cases of fully relaxed and fixed ZnO lattice constant. Efficient spin polarized electron injection is predicted to be achieved at the ZnO/NiO (0001) interface, since the desired half-metallic properties are retained. Under a tough lattice strain, the half-metallic property become slightly weak since values of the majority band gap and spin-flip gap decrease to 0.66 eV and 0.24 eV, respectively. Magnetic configuration of the interface system keeps a ferromagnetic ground state as *w*-NiO does. We also find that the interface effects are highly localized and only make few changes to the properties of bulk ZnO and *w*-NiO materials. As a consequence, the combination of *w*-NiO and ZnO at (0001) interface has promising applications in spintronics.

# Ⅴ. ACKNOWLEDGEMENTS

This research work is supported by the National High-tech R&D Program of China (863 Program) (Grant No：2009AA01A402) and Innovative Foundation of Huazhong University of Science and Technology (Grant No. C2009Q007). Computational resources provided by Wuhan National Laboratory for Optoelectronics (WNLO).